\documentclass[11pt]{article}
\usepackage[utf8]{inputenc}
\usepackage{amsfonts}
\usepackage{amsmath}
\usepackage{amssymb}
\usepackage{indentfirst}
\usepackage{graphicx}
\usepackage[colorlinks]{hyperref}
\usepackage{cite}
\usepackage{array}
\usepackage{microtype}
\usepackage{soul}
\usepackage[normalem]{ulem}
\usepackage{cancel}
\usepackage{soul}
\usepackage{xspace}

\numberwithin{equation}{section}
\allowdisplaybreaks

\makeatletter

\begin{document}

\begin{titlepage}
\vspace{3cm}

\baselineskip=24pt

\begin{center}
\textbf{\LARGE{Non-relativistic gravity theories in four spacetime dimensions}}
\par\end{center}{\LARGE \par}

\begin{center}
	\vspace{1cm}
	\textbf{Patrick Concha}$^{\ast}$,
	\textbf{Evelyn Rodríguez}$^{\ast}$,
	\textbf{Gustavo Rubio}$^{\ast}$,
	\small
	\\[5mm]
    $^{\ast}$\textit{Departamento de Matemática y Física Aplicadas, }\\
	\textit{ Universidad Católica de la Santísima Concepción, }\\
\textit{ Alonso de Ribera 2850, Concepción, Chile.}
	 \\[5mm]
	\footnotesize
	\texttt{patrick.concha@ucsc.cl},
	\texttt{erodriguez@ucsc.cl},
	\texttt{gustavo.rubio@ucsc.cl}
	\par\end{center}
\vskip 26pt
\begin{abstract}

In this work we present a non-relativistic gravity theory defined in four spacetime dimensions using the MacDowell-Mansouri geometrical formulation. We obtain a Newtonian gravity action which is constructed from the curvature of a Newton-Hooke version of the so-called Newtonian algebra. We show that the non-relativistic gravity theory presented here contains the Poisson equation in presence of a cosmological constant. Moreover we make contact with the Modified Newtonian Dynamics (MOND) approach for gravity by considering a particular ansatz for a given gauge field. We extend our results to a generalized non-relativistic MacDowell-Mansouri gravity theory by considering a generalized Newton-Hooke algebra.

\end{abstract}
\end{titlepage}\newpage {} 

{\baselineskip=12pt \tableofcontents{}}

\section{Introduction}

Gravity is successfully described by Einstein's theory of General Relativity providing a geometrical interpretation to the gravitational force, whose underlying (pseudo-)Riemannian geometry ensures local Lorentz symmetry. Interestingly, a geometrical interpretation of Newtonian gravity has been subsequently presented in \cite{Cartan1,Cartan2} through the Newton-Cartan gravity formulation whose underlying geometry is denoted as Newton-Cartan geometry. This formalism allows us to describe the coupling of gravity to massive non-relativistic particles and field theories. There has been a growing interest in non-relativistic theories due to recent applications of Newton-Cartan geometry in strongly coupled condensed matter systems and non-relativistic effective field theories \cite{Son:2008ye,Balasubramanian:2008dm,Kachru:2008yh,Taylor:2008tg,Duval:2008jg,Bagchi:2009my,Hartnoll:2009sz,Bagchi:2009pe,Hoyos:2011ez,Son:2013rqa,Christensen:2013lma,Christensen:2013rfa,Abanov:2014ula,Hartong:2014oma,Hartong:2014pma,Hartong:2015wxa,Geracie:2015dea,Gromov:2015fda,Hartong:2015zia,Taylor:2015glc,Zaanen:2015oix,Hartong:2016yrf,Devecioglu:2018apj}.

The Newton-Cartan geometry allows to geometrize the Poisson equation of Newtonian gravity. However, an action principle for Newtonian gravity, analogous to the Einstein-Hilbert action in General Relativity, has been introduced only recently in \cite{Hansen:2018ofj} by considering a non-relativistic algebra beyond the Bargmann algebra \cite{Grigore:1993fz,Bose:1994sj,Duval:2000xr,Jackiw:2000tz,Papageorgiou:2009zc,Bergshoeff:2016lwr,Concha:2019lhn}. The underlying symmetry, denoted as Newtonian algebra, has then been extended to construct a non-degenerate Chern-Simons gravity action in three spacetime dimensions \cite{Ozdemir:2019tby}. The inclusion of a cosmological constant and its derivation from a relativistic algebra has subsequently been studied in \cite{Concha:2019dqs,Concha:2022you} and \cite{Gomis:2019nih,Bergshoeff:2020fiz}, respectively. Interestingly, the Newton-Hooke version of the Newtonian algebra along its post-Newtonian extensions result to appear from the relativistic $\mathfrak{so}\left(3,2\right)$ algebra considering the Lie algebra expansion method.

Non-relativistic versions of the $\mathfrak{so}\left(3,2\right)$ algebra can be obtained considering the algebra expansion method based on semigroups \cite{Izaurieta:2006zz,Caroca:2011qs,Andrianopoli:2013ooa,Artebani:2016gwh,Ipinza:2016bfc,Penafiel:2016ufo,Inostroza:2018gzd}. The expansion procedure was first introduced in \cite{Hatsuda:2001pp} and subsequently developed in \cite{deAzcarraga:2002xi,deAzcarraga:2007et} in terms of the Maurer-Cartan forms. In particular, the semigroup expansion (S-expansion) method basically consists in combining the multiplication law of the semigroup $S$ with the structure constants of a given Lie algebra $\mathfrak{g}$. Then, the expanded algebra is given by $\mathfrak{G}=S\times\mathfrak{g}$. Remarkably, the S-expansion procedure has two main advantages. First, it provides us with the non-vanishing components of the invariant tensor for the expanded algebra in terms of the original ones which are crucial in the construction of an action. Secondly, it allows us to obtain all the expanded algebras derived with the expansion method based on Maurer-Cartan forms, with a particular semigroup. Thus, the S-expansion method offers us a larger catalogue of expanded algebras from a given one. Interestingly, as it was shown in \cite{Gomis:2019nih,Concha:2020tqx,Concha:2021jos,Concha:2022muu,Concha:2022you}, the $S_{E}^{\left(N\right)}$ semigroup can act as a Galilean expansion by considering a particular subspace decomposition of the original Lie algebra. 

In this paper we present a geometrical formulation, diverse to the Newton-Cartan formalism \cite{Duval:1983pb,Duval:1984cj,Duval:2009vt,Andringa:2010it,Banerjee:2014nja,Banerjee:2016laq,Bergshoeff:2017dqq,Bergshoeff:2022eog}, of non-relativistic gravities in four spacetime dimensions using the MacDowell-Mansouri gravity approach \cite{MacDowell:1977jt}. To this end, we first consider the Newton-Hooke version of the Newtonian algebra \cite{Gomis:2019nih} and use the $S_{E}$-expansion method to derive the respective invariant bilinear trace needed to construct a MacDowell-Mansouri action. The obtained theory describes Newtonian gravity whose equations of motion can be seen as the non-relativistic version of the Einstein equations with a non-vanishing cosmological constant. In presence of matter, the dynamics is given by the Poisson equation in presence of a cosmological constant. Moreover, we show that considering an ansatz for a particular gauge field, we are able to make contact with the Modified Newtonian Dynamics (MOND) approach for gravity \cite{Bekenstein:1984tv,Milgrom:1983M,Milgrom:1983zz}. We end our work with the construction of a non-relativistic gravity action à la MacDowell-Mansouri by considering a generalization of the Newton-Hooke algebra introduced in \cite{Gomis:2019nih}. The generalized obtained theory contains our Newtonian gravity theory as a particular sub-case.

The paper is organized as follows: In section \ref{sec2} we briefly review the relativistic MacDowell-Mansouri gravity theory. Sections \ref{sec3} and \ref{sec4} contain our mains results. In section \ref{sec3} we present the four-dimensional non-relativistic gravity theory using the MacDowell-Mansuri formalism. In section \ref{sec4}, we extend our results to a generalized Newton-Hooke algebra and present the corresponding gravity theory. Section \ref{Ccls} is devoted to some concluding remarks and discussion about future applications of our results.


\section{Four-dimensional gravity theory à la MacDowell-Mansouri}\label{sec2}

A four-dimensional action for gravity can be written considering the MacDowell-Mansouri formalism \cite{MacDowell:1977jt,Wise:2006sm}. Such unified geometric approach is based on the relativistic AdS Lie algebra in four spacetime dimensions given by $\mathfrak{so}\left(3,2\right)$. Remarkably, the MacDowell-Mansouri action reproduces the Einstein gravity with cosmological constant plus boundary terms. In this section, we present a brief review of such geometric construction along with our notation.

The commutation relations for the $\mathfrak{so}\left(3,2\right)$ algebra are given by
\begin{align}
    \left[\hat{J}_{AB},\hat{J}_{CD}\right]&=\eta_{\left[A\left[C\right.\right.}\hat{J}_{\left.\left.D\right]B\right]}\,, \notag \\
    \left[\hat{J}_{AB},\hat{P}_{C}\right]&=\eta_{C\left[B\right.}\hat{P}_{\left.A\right]}\,, \notag \\
    \left[\hat{P}_{A},\hat{P}_{B}\right]&=\frac{1}{\ell^2}\hat{J}_{AB}\,. \label{AdS4}
\end{align}
where $A=0,1,2,3$ are Lorentz indices which are lowered and raised with the Minkowski metric $\eta_{AB}=\left(-1,1,1,1\right)$ and the $\ell$ parameter is related to the cosmological constant through $\Lambda = - \frac{3}{\ell^2}$. One can notice that the vanishing cosmological constant limit $\ell\rightarrow\infty$ leads us to the Poincaré algebra. Here the anti-symmetrization, denoted with brackets $\left[\,\cdots\right]$, is performed without normalization factor, i.e.,
\begin{equation}
    \hat{J}_{\left[AB\right]}=\hat{J}_{AB}-\hat{J}_{BA}\,.
\end{equation}
The gauge connection one-form  $A=A^{a}_{\mu}T_{a}\otimes dx^{\mu}$ for the $\mathfrak{so}\left(3,2\right)$ algebra is given by
\begin{align}
    A&=\frac{1}{2}W^{AB}\hat{J}_{AB}+E^{A}\hat{P}_{A}\,, \label{A}
\end{align}
where $W^{AB}$ is the spin-connection one-form and $E^{A}$ is the vierbein one-form. The corresponding curvature two-form $F=dA+\frac{1}{2}\left[A,A\right]$ reads
\begin{align}
    F&=\frac{1}{2}\mathcal{R}^{AB}\hat{J}_{AB}+T^{A}\hat{P}_{A}\,, \label{F}
\end{align}
with $\mathcal{R}^{AB}$ and $T^{A}$ being the respective AdS curvature and torsion,
\begin{align}
    \mathcal{R}^{AB}&=dW^{AB}+W^{A}_{\ C}W^{CB}+\frac{1}{\ell^2}E^{A}E^{B}\,,\notag \\
    T^{A}&=dE^{A}+W^{A}_{\ B}E^{B}\,.\label{FDA1}
\end{align}
The MacDowell-Mansouri gravity action \cite{MacDowell:1977jt} is constructed with the curvature two-form \eqref{F} and is given by
\begin{align}
    I_{MM}&=2\int_{\mathcal{M}_{4}}\langle F F \rangle\,, \label{MM}
\end{align}
where $\langle\cdots\rangle$ denotes the bilineal invariant trace for the $\mathfrak{so}\left(3,1\right)$ algebra. Let us note that if we consider the invariant tensor for the whole $\mathfrak{so}\left(3,2\right)$ Lie algebra, the action reduces to a topological invariant. Nevertheless, MacDowell and Mansouri realized that considering the bilinear invariant trace for the Lorentz subalgebra $\mathfrak{so}\left(3,1\right)$, given by
\begin{align}
    \langle \hat{J}_{AB}\hat{J}_{CD}\rangle&=\sigma \epsilon_{ABCD}\,, \label{IT1}
\end{align}
it is possible to construct a gravity action. Indeed, considering the invariant tensor \eqref{IT1} and the curvature two-form \eqref{F} in the general expression of the MacDowell-Mansouri action \eqref{MM}, we obtain
\begin{align}
    I_{MM}&=\frac{\sigma}{2}\int_{\mathcal{M}_4} \epsilon_{ABCD}\mathcal{R}^{AB}\mathcal{R}^{CD}\,, \label{MMAdS1}
\end{align}
which can be written considering the explicit components of the curvature two-form \eqref{F} as
\begin{align}
    I_{MM}&=\frac{\sigma}{2}\int_{\mathcal{M}_4} \epsilon_{ABCD}\left(R^{AB}R^{CD}+\frac{2}{\ell^2}R^{AB}E^{C}E^{D}+\frac{1}{\ell^4}E^{A}E^{B}E^{C}E^{D}\right)\,, \label{MMAdS}
\end{align}
where $R^{AB}=dW^{AB}+W^{A}_{\ C}W^{CB}$ is the usual Lorentz curvature two-form. Note that the $\sigma$ constant is related with the gravitational constant. The MacDowell-Mansouri gravity action reproduces the Einstein-Hilbert term, the cosmological constant term and a Gauss-Bonnet term $\epsilon_{ABCD}R^{AB}R^{CD}$ being a boundary term. Such action is equivalent to the four-dimensional Born-Infeld gravity action \cite{Troncoso:1999pk} which can be seen as the Pfaffian of the curvature two-form. One can note that considering the invariant tensor as in \eqref{IT1} breaks the $SO\left(3,2\right)$ group to its Lorentz subgroup. Although the symmetry group is reduced to a Lorentz one, the geometric formalism proposed by MacDowell and Mansouri in \cite{MacDowell:1977jt} allows us to construct a gravity action in four spacetime dimensions whose equations of motion are those of General Relativity with cosmological constant, namely
\begin{align}
    \epsilon_{ABCD}\left(R^{AB}E^{C}+\frac{1}{\ell^2}E^{A}E^{B}E^{C}\right)=0\,, \notag \\
    \epsilon_{ABCD}T^{C}E^{D}=0\,, \label{GReom}
\end{align}
One can note that the vanishing cosmological constant limit $\ell\rightarrow\infty$ leads us to the Einstein equations. However, a MacDowell-Mansouri action based on the curvature two-forms for the $\mathfrak{iso}\left(3,1\right)$ algebra reduces to the Gauss-Bonnet term and the theory does not describe Einstein gravity anymore. Such difficulty can be overcome by considering a larger symmetry as the Maxwell one. As it was shown in \cite{Concha:2013uhq,Concha:2014vka}, General Relativity without cosmological constant appears considering the Born-Infeld gravity action with the Maxwell curvature two-forms. In presence of supersymmetry, the MacDowell-Mansouri action is obtained considering the curvatures of the $\mathfrak{osp}\left(4,1\right)$ superalgebra \cite{MacDowell:1977jt}. Further generalizations of such geometrical approach in the supergravity context have then be developed in \cite{Townsend:1977fz,Castellani:2013iq,Andrianopoli:2014aqa,Concha:2014tca,Concha:2015tla,Penafiel:2018vpe,Concha:2018ywv,Andrianopoli:2020zbl,Eder:2021rgt,Alvarez:2021qbu,Andrianopoli:2021rdk}.

\section{Non-relativistic gravity theory in four spacetime dimensions}\label{sec3}

In this section we present a four-dimensional non-relativistic gravity theory using the MacDowell-Mansouri formalism. We show that a non-relativistic gravity action can be formulated in a geometric way and whose equations of motion correspond to Newton Cartan gravity with cosmological constant. Unlike the standard procedure to derive Newton-Cartan gravity by gauging the Bargmann algebra, here the non-relativistic gravity action is obtained considering uniquely the curvatures of the Newton-Hooke version of the Newtonian algebra introduced in \cite{Hansen:2018ofj}. 
Before approaching the explicit construction of the non-relativistic action à la MacDowell-Mansouri, we first review the non-relativistic expansion of the $\mathfrak{so}\left(3,2\right)$ algebra following \cite{Gomis:2019nih}. Such procedure will be used to obtain the corresponding invariant tensor needed to construct a MacDowell-Mansouri action. To this end, we start by decomposing the relativistic $A$ index in terms of space and time components $A=\left(0,a\right)$ with $a=1,2,3$. With such decomposition, the AdS algebra in four spacetime dimensions reads
\begin{align}
    \left[\hat{J}_{ab},\hat{J}_{cd}\right]&=\delta_{\left[a\left[c\right.\right.}\hat{J}_{\left.\left.d\right]b\right]} \,, &\left[\hat{J}_{ab},\hat{J}_{c}\right]&=\delta_{c\left[b\right.}\hat{J}_{\left.a\right]}\,, \notag \\
    \left[\hat{J}_{a},\hat{J}_{b}\right]&=\hat{J}_{ab} \,, &\left[\hat{J}_{ab},\hat{P}_{c}\right]&=\delta_{c\left[b\right.}\hat{P}_{\left.a\right]}\,, \notag \\ \left[\hat{J}_{a},\hat{P}\right]&=\hat{P}_{a}\,,&\left[\hat{J}_{a},\hat{P}_{b}\right]&=\delta_{ab}\hat{P}\,, \notag \\
    \left[\hat{P}_{a},\hat{P}_{b}\right]&=\frac{1}{\ell^2}\hat{J}_{ab} \,, &\left[\hat{P},\hat{P}_{a}\right]&=\frac{1}{\ell^2}\hat{J}_{a} \,, \label{AdSd}
\end{align}
where we have relabelled the AdS generators as,
\begin{align}
    \hat{J}_{a}&=\hat{J}_{0a}\,, &\hat{J}_{ab}&=\hat{J}_{ab}\,, &\hat{P}&=\hat{P}_{0}\,, &\hat{P}_{a}&=\hat{P}_{a}\,. 
\end{align}
Furthermore, considering the aforementioned decomposition, the invariant tensor \eqref{IT1} can be written as follows
\begin{align}
    \langle \hat{J}_{ab} \hat{J}_{c}\rangle&=\sigma \epsilon_{abc}\,. \label{ITNR}
\end{align}

\subsection{Non-relativistic expansion of the $\mathfrak{so}\left(3,2\right)$ algebra}

As it was shown in \cite{Gomis:2019nih}, a Newton-Hooke version of the Newtonian algebra introduced in \cite{Hansen:2018ofj} can be obtained by considering the $S_{E}^{\left(3\right)}$-expansion to the $\mathfrak{so}\left(3,2\right)$ algebra. To this end, let us first consider the subspace decomposition $\mathfrak{so}\left(3,2\right)=V_0\oplus V_1$ where $V_0=\{\hat{J}_{ab},\hat{P}\}$ and $V_1=\{\hat{J}_{a},\hat{P}_{a}\}$. One can notice that such decomposition satisfies
\begin{align}
[V_0,V_0]&\subset V_0\,, \quad &[V_0,V_1]&\subset V_1\,, \quad &[V_1,V_1]&\subset V_0\,.\label{sd}
\end{align}
Then, let $S_{E}^{\left(3\right)}=S_0\cup S_1$ be a subset decomposition of the semigroup with
\begin{align}
    S_0&=\{\lambda_0,\lambda_2,\lambda_4\}\,, \notag \\
S_1&=\{\lambda_1,\lambda_3,\lambda_4\} \,, 
\label{sd2}
\end{align}
where the elements of the semigroup $S_{E}^{\left(3\right)}$ satisfy the following multiplication law
\begin{equation}
\begin{tabular}{l|lllll}
$\lambda _{4}$ & $\lambda _{4}$ & $\lambda _{4}$ & $\lambda _{4}$ & $\lambda
_{4}$ & $\lambda_{4}$ \\
$\lambda _{3}$ & $\lambda _{3}$ & $\lambda _{4}$ & $\lambda _{4}$ & $\lambda
_{4}$ & $\lambda_{4}$ \\
$\lambda _{2}$ & $\lambda _{2}$ & $\lambda _{3}$ & $\lambda _{4}$ & $\lambda
_{4}$ & $\lambda_{4}$ \\
$\lambda _{1}$ & $\lambda _{1}$ & $\lambda _{2}$ & $\lambda _{3}$ & $\lambda
_{4}$ & $\lambda_{4}$ \\
$\lambda _{0}$ & $\lambda _{0}$ & $\lambda _{1}$ & $\lambda _{2}$ & $\lambda
_{3}$ & $\lambda_{4}$ \\ \hline
& $\lambda _{0}$ & $\lambda _{1}$ & $\lambda _{2}$ & $\lambda _{3}$ & $\lambda_{4}$%
\end{tabular}
\label{mlSE3}
\end{equation}
Here $\lambda_4=0_S$ is the zero element of the semigroup $\lambda_4\lambda_{\alpha} = \lambda_{\alpha} \lambda_{4}=\lambda_{4}$. One can notice that the subset decomposition is said to be resonant since it satisfies the same algebraic structure than the subspaces $V_0$ and $V_1$. One finds a non-relativistic algebra after extracting a resonant subalgebra of the $S_{E}^{\left(3\right)}$-expansion of the $\mathfrak{so}\left(3,2\right)$ algebra,
\begin{align}
    \mathfrak{G}_{R}&=S_0\times V_0\oplus S_{1}\times V_{1}\,,
\end{align}
and considering the $0_S$-reduction conduction $0_S T_{A}=0$. The expanded generators are related to the relativistic $\mathfrak{so}\left(3,2\right)$ ones through the semigroup elements as
\begin{align}
    J_{ab}&=\lambda_{0}\hat{J}_{ab}\,, &H&=\lambda_0\hat{P}\,, &S_{ab}&=\lambda_2 \hat{J}_{ab}\,, &M&=\lambda_2 \hat{P}\,, \notag \\
    G_{a}&=\lambda_1 \hat{J}_{a}\,, &P_{a}&=\lambda_1 \hat{P}_{a}\,, &B_{a}&=\lambda_{3}\hat{J}_{a}\,, &T_{a}&=\lambda_{3}\hat{P}_{a}\,,
\end{align}
and satisfies the following commutation relations:
\begin{align}
    \left[J_{ab},J_{cd}\right]&=\delta_{\left[a\left[c\right.\right.}J_{\left.\left.d\right]b\right]} \,, &\left[J_{ab},P_{c}\right]&=\delta_{c\left[b\right.}P_{\left.a\right]}\,,  & \left[G_{a},G_{b}\right]&=S_{ab} \,, \notag \\
    \left[J_{ab},S_{cd}\right]&=\delta_{\left[a\left[c\right.\right.}S_{\left.\left.d\right]b\right]} \,,& \left[S_{ab},G_{c}\right]&=\delta_{c\left[b\right.}B_{\left.a\right]}\,,  & \left[G_{a},P_{b}\right]&=\delta_{ab}M \,,  \notag \\
    \left[J_{ab},G_{c}\right]&=\delta_{c\left[b\right.}G_{\left.a\right]}\,, & \left[G_{a},H\right]&=P_{a}\,, & \left[P_{a},P_{b}\right]&=\frac{1}{\ell^2}S_{ab} \,,  \notag \\ 
    \left[J_{ab},B_{c}\right]&=\delta_{c\left[b\right.}B_{\left.a\right]}\,,&  \left[G_{a},M\right]&=T_{a}\,, &  \left[H,P_{a}\right]&=\frac{1}{\ell^2}G_{a}\,, \notag \\
    \left[J_{ab},T_{c}\right]&=\delta_{c\left[b\right.}T_{\left.a\right]}\,, & \left[B_{a},H\right]&=T_{a}\,, & \left[H,T_{a}\right]&=\frac{1}{\ell^2}B_{a}\,, \notag \\
    \left[S_{ab},P_{c}\right]&=\delta_{c\left[b\right.}T_{\left.a\right]}\,, & \left[M,P_{a}\right]&=\frac{1}{\ell^2}B_{a}\,, \label{nhNEW}
\end{align}
where we have considered the multiplication law of the semigroup $S_{E}^{\left(3\right)}$ \eqref{mlSE3} along with the commutation relations of the original algebra $\mathfrak{so}\left(3,2\right)$ \eqref{AdSd}. The expanded algebra corresponds to the Newton-Hooke version of the Newtonian algebra which appears as the underlying symmetry for Newtonian gravity \cite{Hansen:2018ofj}. Such algebra differs from the extended Newton-Hooke algebra \cite{Alvarez:2007fw,Papageorgiou:2010ud,Hartong:2016yrf} due to the presence of the additional generators $\{B_{a},T_{a}\}$ which appear as expansion of the set $\{\hat{J}_{a},\hat{P}_{a}\}$. Naturally, one could get the four-dimensional Newton-Hooke algebra considering a smaller semigroup. However, as we shall see, the minimal algebraic structure allowing us to construct a non-relativistic gravity action with cosmological constant considering the MacDowell-Mansouri approach is the algebra \eqref{nhNEW}. Let us note that considering the relativistic $\mathfrak{iso}\left(3,1\right)$ algebra  as the original algebra instead of $\mathfrak{so}\left(3,2\right)$ would reproduce the Newtonian algebra.

\subsection{A MacDowell-Mansouri formulation of non-relativistic gravity theory}

The construction of MacDowell-Mansouri gravity action requires to start with the gauge connection one-form $A$ for the Newton-Hooke-Newtonian ($\mathfrak{nhNewt}$) algebra obtained previously:
\begin{align}
    A&=\frac{1}{2}\omega^{ab}J_{ab}+\tau H+\omega^{a}G_{a}+e^{a}P_{a}+\frac{1}{2}s^{ab}S_{ab}+m M+ b^{a}B_{a}+t^{a}T_{a}\,, \label{AnhNEW}
\end{align}
where $\omega^{ab}$, $\omega^{a}$, $\tau$ and $e^{a}$ are the time and spatial components of the spin-connection and vierbein, respectively. The corresponding curvature two-form $F=F^{A}T_{A}$ reads
\begin{align}
    F=&\frac{1}{2}R^{ab}\left(\omega\right)J_{ab}+R\left(\tau\right)H+R^{a}\left(\omega\right)G_{a}+R^{a}\left(e\right)P_{a}\notag \\
    &+\frac{1}{2}R^{ab}\left(s\right)S_{ab}+R\left(m\right)M+R^{a}\left(b\right)B_{a}+R^{a}\left(t\right) \,, \label{FnhNEW}
\end{align}
where
\begin{align}
    R^{ab}\left(\omega\right)&=d\omega^{ab}+\omega^{a}_{\ c}\omega^{cb}\,,\notag\\
    R\left(\tau\right)&=d\tau\,, \notag \\
    R^{a}\left(\omega\right)&=d\omega^{a}+\omega^{a}_{\ c}\omega^{c}+\frac{1}{\ell^2}\tau e^{a}\,, \notag \\
    R^{a}\left(e\right)&=de^{a}+\omega^{a}_{\ c}e^{c}+\omega^{a}\tau\,, \notag\\
    R^{ab}\left(s\right)&=ds^{ab}+2\omega^{a}_{\ c}s^{cb}+\omega^{a}\omega^{b}+\frac{1}{\ell^2}e^{a}e^{b}\,, \notag \\
    R\left(m\right)&=dm+\omega^{a}e_{a}\,, \notag \\
    R^{a}\left(b\right)&=db^{a}+\omega^{a}_{\ c}b^{c}+s^{a}_{\ c}\omega^{c}+\frac{1}{\ell^2}\tau t^{a}+\frac{1}{\ell^2}me^{a}\,, \notag \\
    R^{a}\left(t\right)&=dt^{a}+\omega^{a}_{\ c}t^{c}+s^{a}_{\ c}e^{c}+\omega^{a}m+b^{a}\tau\,. \label{FDA2}
\end{align}
In the vanishing cosmological constant $\ell\rightarrow\infty$ the curvature two-forms correspond to the Newtonian ones. Considering the Theorem VII.1 of \cite{Izaurieta:2006zz}, one can show that the $\mathfrak{nhNewt}$ algebra \eqref{nhNEW} admits the following non-vanishing components of the invariant tensor
\begin{align}
    \langle J_{ab}G_{c}\rangle&=\alpha  \epsilon_{abc}\,,\notag \\
    \langle J_{ab}B_{c}\rangle&=\beta \epsilon_{abc}\,,\notag \\
    \langle S_{ab}G_{c}\rangle&=\beta \epsilon_{abc}\,,\label{IT2}
\end{align}
which appear as an $S_{E}^{\left(3\right)}$-expansion of the original invariant tensor \eqref{ITNR}. In particular, $\alpha$ and $\beta$ are arbitrary constants related to the $\mathfrak{so}\left(3,1\right)$ constant through the semigroup elements:
\begin{align}
    \alpha&=\lambda_1\sigma\,, &\beta&=\lambda_3\sigma\,. \label{expCC}
\end{align}
Let us note that the invariant tensor \eqref{IT2} breaks the symmetry to an extended Nappi-Witten subalgebra spanned by $\{J_{ab},G_{a},S_{ab},B_{a}\}$. A four-dimensional non-relativistic gravity action is obtained considering the invariant tensor \eqref{IT2} and the curvature two-forms \eqref{FnhNEW} in the general expression of the MacDowell-Mansouri action \eqref{MM},
\begin{align}
    I_{MM}^{\mathfrak{nhNewt}}&=2\int_{\mathcal{M}_4}\left(\mathcal{L}_{\alpha}+\mathcal{L}_{\beta}\right)\,,
\end{align}
with
\begin{align}
    \mathcal{L}_{\alpha}&=\alpha \,\epsilon_{abc} R^{ab}\left(\omega\right)R^{c}\left(\omega\right)\,, \notag \\
    \mathcal{L}_{\beta}&=\beta\,\epsilon_{abc} \left[R^{ab}\left(\omega\right) R^{c}\left(b\right)+R^{ab}\left(s\right)R^{c}\left(\omega\right)\right]\,.
\end{align}
Then, the non-relativistic MacDowell-Mansouri gravity action can be rewritten by taking into account the explicit expression of the curvature two-forms \eqref{FDA2}:
\begin{align}
    I_{MM}^{\mathfrak{nhNewt}}= &\, 2\int_{\mathcal{M}_4}\frac{\alpha}{\ell^2}\,\epsilon_{abc}\ R^{ab}\left(\omega\right)\tau e^{c} \notag\\
    &+\frac{\beta}{\ell^2}\epsilon_{abc}\left[\mathcal{R}^{a}\left(\omega\right)e^{b}e^{c}+\frac{1}{\ell^2}e^{a}e^{b}\tau e^{c}+\mathcal{R}^{ab}\left(s\right)\tau e^{c}+R^{ab}\left(\omega\right)\tau t^{c}+R^{ab}\left(\omega\right)m e^{c}\right]\,, \label{nrMM}
\end{align}
where we have omitted the boundary terms and considered the redefinition
\begin{align}
    \mathcal{R}^{a}\left(\omega\right)&=d\omega^{a}+\omega^{a}_{\ c}\omega^{c}\,, \notag \\
    \mathcal{R}^{ab}\left(s\right)&=ds^{ab}+2\omega^{a}_{\ c}s^{cb}+\omega^{a}\omega^{b}\,.
\end{align}
The MacDowell-Mansouri non-relativistic gravity action contains two independent sectors proportional to $\alpha$ and $\beta$, respectively. The term proportional to $\alpha$ describes Galilei gravity being the non-relativistic counterpart of Einstein gravity. On the other hand, the term along $\beta$ reproduces a Newtonian gravity in presence of a cosmological constant term where the two first term can be seen as the Newtonian counterpart of General Relativity with cosmological constant. The other terms contains the explicit presence of the extra gauge field $s^{ab}$, $t^{a}$ and $m$. Such additional content is required in order to construct a non-relativistic MacDowell-Mansouri action different to Galilei gravity. Indeed if we would consider the extended Newton-Hooke symmetry, the invariant tensor would only contain the term proportional to $\alpha$. Such particularity can be seen from the S-expansion procedure in which the $\beta$ constant of the invariant tensor is related to the relativistic one through $\lambda_3$. Then, the minimal semigroup having $\lambda_3\neq 0$ is the $S_{E}^{\left(3\right)}$ semigroup allowing us to obtain the $\mathfrak{nhNewt}$ algebra. 

Although the invariant tensor \eqref{IT2} breaks the $\mathfrak{nhNewt}$ algebra to its extended Nappi-Witten subalgebra $\{J_{ab},G_{a},S_{ab},B_{a}\}$, the equations of motion can be seen as the non-relativistic counterpart of the Einstein equations with cosmological constant. Indeed, varying the action with respect to the diverse gauge fields provides us with the following field equations:
\begin{align}
    \delta \omega^{ab}&:\ \alpha\epsilon_{abc}\ \left(d\tau e^{c}-\tau T^{c}\right)+\beta\epsilon_{abc}\ \left(d\tau t ^{c}-\tau \mathcal{R}^{c}\left(t\right)+R\left(m\right)e^{c}-mT^{c}\right)=0\,, \notag \\
    \delta \omega^{a}&:\ \beta\epsilon_{abc}\ R^{b}\left(e\right)e^{c}=0\,, \notag \\
    \delta \tau &: \ \alpha\epsilon_{abc}\ R^{ab}\left(\omega\right)e^{c}+\beta\epsilon_{abc}\ \left(R^{ab}\left(s\right)e^{c}+R^{ab}\left(\omega\right)t^{c}\right)=0 \,,\notag \\
    \delta e^{c}&: \ \alpha\epsilon_{abc}\ R^{ab}\left(\omega\right)\tau+\beta\epsilon_{abc}\left(2R^{a}\left(\omega\right)e^{b}+R^{ab}\left(s\right)\tau+R^{ab}\left(\omega\right)m\right)=0\,, \notag \\
    \delta s^{ab} &: \ \beta\epsilon_{abc}\ \left(d\tau e^{c}-\tau T^{c}\right)=0 \,, \notag \\
    \delta m &: \ \beta\epsilon_{abc}\ R^{ab}\left(\omega\right)e^{c}=0\,, \notag \\
    \delta t^{c} &: \ \beta\epsilon_{abc}\ R^{ab}\left(\omega\right)\tau=0\,, \label{eom1}
\end{align}
where 
\begin{align}
    T^{a}&=de^{a}+\omega^{a}_{c}e^{c}\,, \notag \\ \mathcal{R}^{a}\left(t\right)&=dt^{a}+\omega^{a}_{\ c}t^{c}+s^{a}_{\ e}e^{c}\,.
\end{align}
The variation of the action \eqref{nrMM} under gauge symmetry is obtained by considering $\delta F=\left[\varepsilon,F\right]$ with $\varepsilon$ being the gauge parameter. In particular, considering the variation of the action under an arbitrary spatial spin-connection $\omega^{a}$ reproduces the $\omega^{a}$ field equation. Then, demanding invariance of the action for arbitrary $\delta \omega^{a}$ yields $R^{a}\left(e\right)=0$ allowing to express the spatial and time spin-connection in terms of the spatial and time vierbein. Let us note that there is no dependence on an extra gauge field corresponding to the central charge generator, as in the Newton-Cartan gravity theory \cite{Andringa:2010it}. Moreover, no curvature constraints have been imposed to solve the spin-connection.

Let us note that the non-relativistic MacDowell-Mansouri gravity action \eqref{nrMM} can be alternatively obtained from the usual relativistic MacDowell-Mansouri action \eqref{MMAdS} by considering the expansion at the level of the gauge fields. Indeed, the non-relativistic gauge fields \eqref{AnhNEW} can be written in terms of the $\mathfrak{so}\left(3,2\right)$ ones through the semigroup elements as
\begin{align}
    \omega^{ab}&=\lambda_0 W^{ab}\,, &\tau&=\lambda_0 E^{0}\,, &s^{ab}&=\lambda_2 W^{ab}\,, &m&=\lambda_2E^{0}\,, \notag \\
    \omega^{a}&=\lambda_1W^{0a}\,, &e^{a}&=\lambda_1 E^{a}\,, &b^{a}&=\lambda_3 W^{0a}\,, &t^{a}&=\lambda_3 E^{a}\,. \label{expGF}
\end{align}
Then, the four-dimensional non-relativistic gravity action \eqref{nrMM} is obtained from the relativistic one considering the expanded gauge fields \eqref{expGF} along the expanded coupling constants \eqref{expCC}.

Although the $S_{E}^{(3)}$-expansion of the $\mathfrak{so}(3,2)$ algebra allows us the construction of a finite NR gravity action, it suffers from a degeneracy problem. Indeed, the Newton-Hooke version of the Newtonian algebra \eqref{nhNEW} only allows for a degenerate invariant tensor.

\subsection{The Poisson equation and the Modified Newtonian Dynamic approach}
In presence of matter, the total Lagrangian corresponds to $\mathcal{L}_{T}=\mathcal{L}^\mathfrak{nhNewt}
_{MM}+\mathcal{L}_{M}$, where $\mathcal{L}_{M}$ represents the matter Lagrangian. Then, the variation of the action with respect to the time vierbein $\tau$ allow us to obtain the Poisson equation and certain post Newtonian generalizations when we consider a perfect fluid without pression. In this context, the non-vanishing component of the energy-momemtum tensor is proportional to the matter density $\rho$ i.e $\mathcal{T}_{00}=\rho$.  Thus, in terms of the matter Lagrangian, the one-form energy-momemtum $\mathcal{T}_{0}$ is given by
\begin{align}
    \mathcal{T}_{0}=\star\frac{\delta\mathcal{L}_{M}}{\delta\tau}\,, \notag
\end{align}
where $\star$ is the Hodge star operator. One can notice that the variation of the action for arbitrary $\delta m$ and $\delta t^a$ yields to $R^{ab}(\omega)=0$  which satisfies the Trautman and Ehlers conditions \cite{Trautman,Ehlers}. The vanishing of the curvature of spatial rotations \cite{Andringa:2010it} implies that the spatial hypersurfaces are flat, i.e. one can choose a coordinate frame with null spatial  $\Gamma$-connection and $\omega_{ab}$ depends only on time. Thus, we can note from the geodesic equation that the only non-zero component of the $\Gamma$-connection is $\Gamma^a_{00}=\delta^{ab}\partial_b\phi(x)$, where $\phi(x)$ is the time independent gravitational potential. One can obtain, after considering the subtraction of the variations with respect to $\tau$ and $e^a$ gauge fields and computing the trace, the following equation in tensorial language
\begin{align}
   \nabla ^{2}\phi(x)=D_a\omega^a_0=-\frac{\ell^2}{4\beta}\rho+\frac{6}{\ell^2}\,, \label{poisson1}
\end{align}
where $\omega^a=\omega^a_B dx^B$ and $\omega^a_B=-\tau_B\delta^{ac}\partial_c\phi(x)$, while the covariant derivative is given by $D=\partial+\Gamma$. In order to make contact with the Poisson equation in the flat limit $\ell\rightarrow\infty$ from (\ref{poisson1}), we require that the arbitrary constant satisfies 
\begin{align}
\beta=-\frac{\ell^2}{16\pi G}\,,\label{beta}
\end{align}
where $G$ is the universal gravitational constant. Interestingly, it is possible to obtain the MOND approach for gravity \cite{Bekenstein:1984tv,Milgrom:1983M,Milgrom:1983zz} from the variation of the action for arbitrary $\delta \tau$. To this end, we consider the following ansatz for the non-zero components of the field $s^{ab}=s^{ab}_Cdx^C$
\begin{equation*}
s_{\text{ \ }0}^{a0}=-s_{\text{ \ }0}^{0a}=\left(\mu(x)+1\right)\delta ^{ab}\partial_{b}\varphi (x)\,,
\end{equation*}
where $\mu (x)$ is the arbitrary function coming from MOND scheme. The function $\varphi(x)$ depends of the gravitational potential $\phi(x)$ and the cosmological constant in the following way
\begin{equation*}
\varphi (x)=\phi(x)-\frac{9}{2\ell^2}\delta_{ab}x^ax^b\,,
\end{equation*}
One can show that $\varphi\left(x\right)$ satisfies the MOND expression \cite{Bekenstein:1984tv} in presence of a cosmological constant:
\begin{equation}
\mu (x)\nabla ^{2}\varphi\left(x\right) =4\pi G\rho -\frac{3}{\ell^2} -\vec{\nabla}\mu (x)\cdot \vec{\nabla}%
\varphi (x)\,,\label{mond}
\end{equation}
where we have considered $\beta$ as in \eqref{beta}. In particular, the Poisson equation \eqref{poisson1} with $\beta=-\frac{\ell^2}{16\pi G}$ is recovered considering $\mu (x)=1$ and can be written in terms of the potential $\varphi (x)$ as
\begin{equation}
\nabla^{2}\varphi (x)=4\pi G\rho-\frac{3}{\ell^2}\,. \label{poisson2}
\end{equation}
\section{Four-dimensional non-relativistic gravity theories and generalized Newton-Hooke symmetries}\label{sec4}
In this section, we consider a generalization of the previous section by extending our construction to a non-relativistic gravity theory using the curvature two-form for an generalized version of the Newton-Hooke symmetry, introduced in \cite{Gomis:2019nih} through the semigroup expansion method. Indeed, an generalized extension of the Newton-Hooke algebra can be obtained considering an bigger semigroup $S_{E}^{\left(N\right)}$, whose elements satisfy
\begin{equation}
\lambda _{\alpha }\lambda _{\beta }=\left\{ 
\begin{array}{lcl}
\lambda _{\alpha +\beta }\,\,\,\, & \mathrm{if}\,\,\,\,\alpha +\beta \leq
N+1\,, &  \\ 
\lambda _{N+1}\,\,\, & \mathrm{if}\,\,\,\,\alpha +\beta >N+1\,, & 
\end{array}%
\right.   \label{mlSEN}
\end{equation}%
where $\lambda_{N+1}=0_S$ is the zero element. Let us consider now a subset decomposition of the semigroup $S_{E}^{\left(N\right)}=S_{0}\cup S_{1}$ with
\begin{eqnarray}
S_{0} &=&\left\{ \lambda _{2m},\ \text{with }m=0,\ldots ,\left[ \frac{N}{2}%
\right] \right\} \cup \{\lambda _{N+1}\}\,, \notag \\
S_{1} &=&\left\{ \lambda _{2m+1},\ \text{with }m=0,\ldots ,\left[ \frac{N-1}{%
2}\right] \right\} \cup \{\lambda _{N+1}\}\,, \label{sdN}
\end{eqnarray}%
where $[\ldots ]$ denotes the integer part.  Note that the subset decomposition \eqref{sdN} is resonant since it satisfies the same algebraic structure than the subspace decomposition \eqref{sd}. Then, an expanded algebra is obtained after considering a resonant $S_{E}^{\left(N\right)}$-expansion and applying a $0_S$-reduction to the $\mathfrak{so}\left(3,2\right)$ algebra. The expanded generators are related to the relativistic AdS ones through the semigroup elements as follows
\begin{align}
    \lambda_{2m}\hat{J}_{ab}& = J_{ab}^{(m)}\,, & \lambda_{2m+1}\hat{J}_{a} &=J_{a}^{(m)}\,, \notag\\
    \lambda_{2m}\hat{P}& = P^{(m)}\,, & \lambda_{2m+1}\hat{P}_{a} &=P_{a}^{(m)}\,.
\end{align}
Then, considering the multiplication law of the semigroup $S_{E}^{\left(N\right)}$ \eqref{mlSEN} and the original commutation relations of the AdS algebra \eqref{AdSd}, it is not difficult to show that the expanded generators satisfy the following commutation relations
\begin{align}
    \left[J_{ab}^{(m)},J_{cd}^{(n)}\right]&=\delta_{\left[a\left[c\right.\right.}J_{\left.\left.d\right]b\right]}^{(m+n)} \,, &\left[J_{ab}^{(m)},J_{c}^{(n)}\right]&=\delta_{c\left[b\right.}J_{\left.a\right]}^{(m+n)}\,, \notag \\
    \left[J_{a}^{(m)},J_{b}^{(n)}\right]&=J_{ab}^{(m+n+1)} \,, &\left[J_{ab}^{(m)},P_{c}^{(n)}\right]&=\delta_{c\left[b\right.}P_{\left.a\right]}^{(m+n)}\,, \notag \\ \left[J_{a}^{(m)},P^{(n)}\right]&=P_{a}^{(m+n)}\,,&\left[J_{a}^{(m)},P_{b}^{(n)}\right]&=\delta_{ab}P^{(m+n+1)}\,, \notag \\
    \left[P_{a}^{(m)},P_{b}^{(n)}\right]&=\frac{1}{\ell^2}J_{ab}^{(m+n+1)} \,, &\left[P^{(m)},P_{a}^{(n)}\right]&=\frac{1}{\ell^2}J_{a}^{(m+n)} \,. \label{infNH}
\end{align}
This non-relativistic algebra can be seen as an infinite-dimensional extension of the Newton-Hooke algebra \cite{Gomis:2019nih} for $N=\infty$ which, in the vanishing cosmological constant limit $\ell\rightarrow\infty$, leads to the infinite-dimensional extension of the Galilei algebra in four dimensions \cite{Hansen:2019vqf,Gomis:2019fdh}. Let us note that for $N=1$, the algebra reproduces the Newton-Hooke algebra \cite{Aldrovandi:1998im,Gibbons:2003rv,Brugues:2006yd,Duval:2011mi,Duval:2016tzi}, whose flat limit corresponds to the Galilei symmetry.  The case with $N=2$ leads to an extended Newton-Hooke algebra \cite{Alvarez:2007fw,Papageorgiou:2010ud,Hartong:2016yrf} which in the flat limit reproduces the extended Bargmann algebra \cite{Bergshoeff:2016lwr}. For $N=3$, we obtain the Newton-Hooke version of the Newtonian algebra considered in the previous section to construct a four-dimensional non-relativistic MacDowell-Mansouri gravity action. An enhanced Bargmann-Newton-Hooke algebra \cite{Bergshoeff:2020fiz}\footnote{also denoted as exotic Newtonian algebra \cite{Concha:2019dqs} or Newton-Hooke version of the extended Post-Newtonian algebra \cite{Gomis:2019nih}} appears for $N=4$ which in the vanishing cosmological constant limit leads to the extended Newtonian symmetry used to construct a Chern-Simons Newtonian gravity action \cite{Ozdemir:2019orp}. 

Let us consider now the construction of a four-dimensional non-relativistic gravity action considering the curvature two-form for the generalized Newton-Hooke algebra \eqref{infNH}, which we shall denote as the $\mathfrak{gnh}^{\left(N\right)}$ algebra. Such symmetry admits the following  non-vanishing components of the invariant tensor
\begin{align}
    \langle J_{ab}^{(m)} J_{c}^{(n)} \rangle&= \alpha_{m+n}\epsilon_{abc}\,,\label{ITEX}
\end{align}
where the arbitrary constants are defined in terms of the  $\mathfrak{so}\left(3,1\right)$ one, $\sigma$, through the semigroup elements as follows
\begin{equation}
    \alpha_{m+n}\equiv \lambda_{2\left(m+n\right)+1}\sigma \,.\label{ccinf}
\end{equation}
Let $A$ be the the gauge connection one-form taking values in the aforesaid algebra:
\begin{equation}
    A=\sum_{m=0}^{\left[\frac{N}{2}\right]}\left(\frac{1}{2}\omega^{ab(m)}{J}^{(m)}_{ab}+\tau^{(m)}P^{(m)}\right)
    +\sum_{m=0}^{\left[\frac{N-1}{2}\right]}\left(\omega^{a(m)}J_{a}^{(m)}+e^{a(m)}P_{a}^{(m)}\right)\,, \label{infA}
\end{equation}
where $\{\omega^{ab(m)},\omega^{(m)},\tau^{(m)},e^{a(m)}\}$ are the expansions of the time and spatial spin-connections and vielbein. The corresponding curvature two-form $F$ reads
\begin{align}
    F=\frac{1}{2}R^{ab}\left(\omega^{(m)}\right){J}^{(m)}_{ab}+R\left(\tau^{(m)}\right)P^{(m)}+R^{a}\left(\omega^{(m)}\right){J}^{(m)}_{a}+R^{a}\left(e^{(m)}\right)P^{(m)}_{a}\,,\label{curvex}
\end{align}
where the components are defined as
\begin{align}
    R^{ab}\left(\omega^{(m)}\right)&= d\omega^{ab(m)}+\sum_{n,l=0}^{\left[\frac{N}{2}\right]}\omega^{a (n)}_{\ c}\omega^{cb (l)}\delta_{n+l}^{m}+\sum_{n,l=0}^{\left[\frac{N-1}{2}\right]}\left(\omega^{a (n)}\omega^{b (l)}+\frac{1}{\ell^2}e^{a (n)}e^{b (l)}\right)\delta_{n+l+1}^{m}\,, \notag\\
    R\left(\tau^{(m)}\right)&=d\tau^{(m)}+\sum_{n,l=0}^{\left[\frac{N-1}{2}\right]}\omega^{a (n)}e_{a}^{(l)}\delta_{n+l+1}^{m}\,, \notag \\
    R^{a}\left(\omega^{(m)}\right)&=d\omega^{a(m)}+\sum_{n,l=0}^{\left[\frac{N}{2}\right]}\left(\omega^{a (n)}_{\ c}\omega^{c(l)}+\frac{1}{\ell^2}\tau^{(n)} e^{a (l)}\right)\delta_{n+l}^{m}\,, \notag \\
    R^{a}\left(e^{(m)}\right)&=de^{(m)}+\sum_{n,l=0}^{\left[\frac{N}{2}\right]}\left(\omega^{a (n)}_{\ c}e^{c (l)}+\omega^{a (n)}\tau^{(l)} \right)\delta_{n+l}^{m}\,.\label{compcurv}
\end{align}
Let us note that, as it is expected, the vanishing cosmological constant limit $\ell\rightarrow\infty$ of \eqref{compcurv} leads to the curvatures of a generalized Galilei symmetry in four spacetime dimensions.  

Then, the four-dimensional non-relativistic gravity action for the $\mathfrak{gnh}^{\left(N\right)}$ algebra can be constructed from the curvature two-form \eqref{curvex} and the invariant tensor \eqref{ITEX} by replacing them in the MacDowell-Mansouri gravity action \eqref{MM}.  The action can be written as follows
\begin{align}
    I_{MM}^{\mathfrak{gnh}^{(N)}}=2\int_{\mathcal{M}_{4}}\mathcal{L}_{\alpha_{i}}\,, \label{INFMM}
\end{align}
where
\begin{align}
    \mathcal{L}_{\alpha_{i}}=\sum_{i=0}^{\left[\frac{N}{2}\right]}\alpha_{i}\,\epsilon_{abc}\,R^{ab}\left(\omega^{(m)}\right)R^{c}\left(\omega^{(n)}\right)\delta_{m+n}^{i}\,.
\end{align}
The non-relativistic MacDowell-Mansouri action can be rewritten considering the explicit expression of the curvature two-forms \eqref{compcurv} as
\begin{align}
    I_{MM}^{\mathfrak{gnh}^{(N)}}&=2\int_{\mathcal{M}_{4}}\sum_{i=0}^{\left[\frac{N}{2}\right]}\alpha_{i}\,\epsilon_{abc}\,\left[\mathcal{R}^{ab}\left(\omega^{\left(m\right)}\right)\tau^{\left(n\right)}e^{c\left(l\right)}\delta^{i}_{m+n+l}+\mathcal{R}^{a}\left(\omega^{\left(m\right)}\right)e^{b\left(n\right)}e^{c\left(l\right)}\delta^{i}_{m+n+l+1}\right.\notag\\
    &\left.+\frac{1}{\ell^2}e^{a\left(m\right)}e^{b\left(n\right)}\tau^{\left(l\right)}e^{c\left(p\right)}\delta^{i}_{m+n+l+p+1} \right] \,. \label{IMMinf}
\end{align}
where we have omitted boundary terms and considered the following redefinition:
\begin{align}
    \mathcal{R}^{ab}\left(\omega^{\left(m\right)}\right)&=d\omega^{ab(m)}+\sum_{n,l=0}^{\left[\frac{N}{2}\right]}\omega^{a (n)}_{\ c}\omega^{cb (l)}\delta_{n+l}^{m}+\sum_{n,l=0}^{\left[\frac{N-1}{2}\right]}\omega^{a (n)}\omega^{b (l)}\delta^{m}_{n+l+1}\,, \notag \\
    \mathcal{R}^{a}\left(\omega^{\left(m\right)}\right)&=d\omega^{a(m)}+\sum_{n,l=0}^{\left[\frac{N}{2}\right]}\omega^{a (n)}_{\ c}\omega^{c(l)}\delta^{m}_{n+l}\,.
\end{align}
One can notice that $\alpha_0$ and $\alpha_1$ correspond to the coupling constants of the non-relativistic gravity action constructed in section \ref{sec3}, namely $\alpha$ and $\beta$, respectively. Indeed, the two first terms of the action \eqref{IMMinf} reproduce the MacDowell-Mansouri gravity action \eqref{nrMM} constructed with the curvature two-forms for the $\mathfrak{nhNewt}$ algebra. Let us note that the general non-relativistic MacDowell-Mansouri gravity action \eqref{INFMM} is modified only for odd values of $N$. For instance, it would require $S_{E}^{\left(5\right)}$ as the relevant semigroup in order to add new contributions to the non-relativistic MacDowell-Mansouri action \eqref{nrMM}. 

On the other hand, it is important to mention that the resulting action \eqref{IMMinf} can be extended to $N=\infty$ and is invariant under the subalgebra spanned by generators ${\{J^{(m)}_{ab},J^{(m)}_{a}\}}$. As in the previous case, considering the invariant tensor for the whole infinite-dimensional Newton-Hooke symmetry would reproduce a topological invariant. The non-relativistic versions of the MacDowell-Mansouri gravity action is obtained considering the invariant tensor as in \eqref{ITEX} which breaks the infinite-dimensional Newton-Hooke symmetry to an infinite-dimensional extension of the Nappi-Witten symmetry \cite{Gomis:2019nih,Concha:2022you}. Interestingly, for arbitrary $N$, the equations of motion can be seen as non-relativistic extensions of the Einstein equations with cosmological constant, namely
\begin{align}
    \delta \tau^{\left(j\right)}&:\ \sum_{i=0}^{\left[\frac{N}{2}\right]}\alpha_{i}\epsilon_{abc}\, \left(\mathcal{R}^{ab}\left(\omega^{\left(m\right)}\right)e^{c\left(n\right)}\delta^{i}_{m+n+j}+\frac{1}{\ell^2}e^{a\left(m\right)}e^{b\left(n\right)}e^{c\left(l\right)}\delta^{i}_{m+n+l+j+1} \right)=0\,,\notag \\
    \delta e^{c\left(j\right)}&:\ \sum_{i=0}^{\left[\frac{N}{2}\right]}\alpha_{i}\epsilon_{abc}\,\left(\mathcal{R}^{ab}\left(\omega^{\left(m\right)}\right)\tau^{\left(n\right)}\delta^{i}_{m+n+j}+2\mathcal{R}^{a}\left(\omega^{\left(m\right)}\right)e^{b\left(n\right)}\delta^{i}_{m+n+j+1}\right.\notag\\
    &\left.+\frac{3}{\ell^2} e^{a\left(m\right)}e^{b\left(n\right)}\tau^{\left(l\right)} \delta^{i}_{m+n+l+j+1}\right) =0\,, \notag \\
    \delta \omega^{a\left(j\right)}&:\ \sum_{i=0}^{\left[\frac{N}{2}\right]}\alpha_{i}\epsilon_{abc}\,\left(R^{b}\left(e^{\left(m\right)}\right)e^{c\left(n\right)} \right)\delta^{i}_{m+n+j+1} =0\,, \notag \\
    \delta \omega^{ab\left(j\right)}&:\ \sum_{i=0}^{\left[\frac{N}{2}\right]}\alpha_{i}\epsilon_{abc}\,\left(R\left(\tau^{\left(m\right)}\right)e^{c\left(n\right)}-\tau^{\left(m\right)}\mathcal{R}^{c}\left(e^{\left(n\right)}\right) \right)\delta^{i}_{m+n+j} =0\,, \label{INFEOM}
\end{align}
where we have defined
\begin{align}
    \mathcal{R}^{a}\left(e^{\left(m\right)}\right)&=de^{(m)}+\sum_{n,l=0}^{\left[\frac{N}{2}\right]}\omega^{a (n)}_{\ c}e^{c (l)}\delta_{n+l}^{m}\,.
\end{align}
Then, demanding invariance of the generalized non-relativistic MacDowell-Mansouri gravity action for arbitrary $\delta\omega^{a\left(m\right)}$ requires to impose the vanishing of the expanded spatial torsion $R^{a}\left(e^{\left(n\right)}\right)=0$ allowing us to express the expanded spatial and time spin connection in terms of the expanded spatial and time vierbein.

As a final remark, let us note that the non-relativistic MacDowell-Mansouri gravity action \eqref{IMMinf} for the generalized Newton-Hooke algebra can also be obtained from the MacDowell-Mansouri action \eqref{MMAdS} by applying the $S$-expansion directly at the level of the relativistic gauge fields. Indeed, the non-relativistic gauge fields appearing in \eqref{infA} can be written in terms of the $\mathfrak{so}\left(3,2\right)$ ones through the semigroup elements as follows
\begin{align}
    \omega^{ab (m)}&=\lambda_{2m} W^{ab}\,, &\tau^{(m)}&=\lambda_{2m} E^{0}\,, \notag \\
    \omega^{a (m)}&=\lambda_{2m+1}W^{0a}\,, &e^{a (m)}&=\lambda_{2m+1} E^{a}\,. \label{expGFinf}
\end{align}
Then, non-relativistic gravity action \eqref{IMMinf} is obtained from the relativistic one considering the expanded gauge fields \eqref{expGFinf} along with the coupling constants defined in \eqref{ccinf}.


\section{Conclusions}\label{Ccls}

In this paper we present a geometrical formulation of non-relativistic gravity using the MacDowell-Mansouri formalism \cite{MacDowell:1977jt} for a Newton-Hooke version of the Newtonian algebra. We show that the obtained Newtonian gravity theory, although it is invariant under only an extended Nappi-Witten subalgebra, reproduces the non-relativistic counterpart of the Einstein equations. Moreover, the invariance of the action requires the vanishing of the spatial torsion allowing to solve the spin-connection without imposing curvature constraints. In presence of matter, we recover not only the Poisson equation for the gravitational potential $\phi(x)$ but also we make contact with the MOND approach for gravity considering a ansatz for the $s^{ab}$ gauge field and by fixing the coupling constant $\beta$. The extension of our results to the generalized Newton-Hooke algebra has also be considered which contains our main result as a particular sub-case.

Our results serves as a starting point for various further developments and studies. It would be interesting to analyze the physical implications of the non-relativistic MacDowell-Mansouri gravity action based on the $\mathfrak{gnh}^{\left(N\right)}$ algebra. For instance, it would be interesting to consider the $N=5$ case which is the minimal symmetry allowing us to extend our non-relativistic MacDowell-Mansouri gravity action. One could explore if post-Newtonian corrections \cite{Dautcourt:1996pm,VandenBleeken:2017rij,Hansen:2019svu,Hansen:2020wqw} or generalizations of the MOND approach can be recovered from such extension. Although the MOND scheme, which involves a modification of the Poisson equation, is known to describe the movement of galactic objects where the asymptotic circular velocity is determined only by the total mass of the galaxy without requiring dark matter, it fails to describe various observed properties of galaxy clusters \cite{Aguirre:2001fj}.

Another point which it would be worth it to study is the flat limit. In absence of a cosmological constant, a MacDowell-Mansouri action built from the curvature of the Newtonian algebra reduces to topological boundary terms. It would be interesting to study if an enlarged non-relativistic structure allows us to define Newtonian gravity (without cosmological constant) à la MacDowell-Mansouri. One could expect that a Newtonian version of the Maxwell algebra \cite{Concha:2020ebl} could be a good candidate to achieve such task since the Maxwell symmetry allows to obtain General Relativity without cosmological constant considering the Born-Infeld gravity action \cite{Concha:2013uhq}. The additional gauge field appearing in the Maxwell symmetry \cite{Schrader:1972zd,Bacry:1970ye,Gomis:2017cmt} have no dynamical contribution and appears on the boundary. One could expect a similar scenario in the non-relativistic counterpart. In such case, one could expect that the novel theory reproduces the Poisson equation and the MOND expression without cosmological constant [work in progress].

A natural continuation of our work is the introduction of supersymmetry in our model. Such generalization remains a challenging task since the non-relativistic limit of a supergravity theory is not trivial. One could follow the method employed in \cite{Concha:2020tqx,Concha:2021jos} which requires to start from a $\mathcal{N}=2$ extended superalgebra. Although a supersymmetric extension of the Newtonian and Newton-Hooke algebras can be obtained through the $S$-expansion procedure, the construction of a non-relativistic supergravity action in four spacetime dimensions using the MacDowell-Mansouri approach requires a subtle treatment as in the relativistic case \cite{Townsend:1977fz,Alvarez:2021qbu}.

\section*{Acknowledgments}

This work was funded by the National Agency for Research and Development ANID - PAI grant No. 77190078, ANID - SIA grant No. SA77210097 and FONDECYT grants No. 1211077, 11220328 and 11220486.  The authors would like to thank to the Dirección de Investigación and Vice-rectoría de Investigación of the Universidad Católica de la Santísima Concepción, Chile, for their constant support.


\bibliographystyle{fullsort.bst}
 
\bibliography{Non_relativistic_MM}

\end{document}